\documentclass[preprint,proceedings]{rmaa}

\suppressfulladdresses 

\usepackage{paralist}

\usepackage{psfrag,color}

\SetYear{2007}
\SetConfTitle{The Nuclear Region, Host Galaxy and Environment of Active Galaxies}

\title{FERO (Finding Extreme Relativistic Objects): statistics of relativistic broad Fe 
K$\alpha$ lines in AGN} 

\author{
  A.L. Longinotti,\altaffilmark{1} 
  I. de La Calle,\altaffilmark{1,2}
  S.Bianchi,\altaffilmark{3}
  M. Guainazzi,\altaffilmark{1} 
  and M. Dov{\v c}iak\altaffilmark{4}}

\altaffiltext{1}{ESAC, P.O. Box 78, 28691 Villanueva de la Ca\~{n}ada, Madrid, Spain.}

\altaffiltext{2}{INSA-ESAC, P.O. Box 78, 28691 Villanueva de la Ca\~{n}ada, Madrid, Spain.}

\altaffiltext{3}{Universit\`a degli Studi di Roma Tre, Roma, Italy}
\altaffiltext{4}{Astronomical Institute, Academy of Sciences, Prague, Czech Republic}

\shortauthor{Longinotti et al.}
\shorttitle{RevMexAA(SC) Demo Document}

\listofauthors{A.L. Longinotti, I.de La Calle, S. Bianchi, M.Guainazzi \and M. Dov{\v c}iak}
\indexauthor{Longinotti A.L.}
\indexauthor{de La Calle I.}
\indexauthor{Bianchi S.}
\indexauthor{Guainazzi M.}
\indexauthor{Dov{\v c}iak M.}

\abstract{The properties of the relativistically broadened Fe K$\alpha$ line emitted in Active Galactic Nuclei (AGN) are still  debated among the AGN community. Recent works seem to exclude that the broad Fe line is a common feature of AGN. 
The analysis of a large sample composed by 157 {\it XMM-Newton} archival observations of radio quiet AGN is presented here. This ongoing project is a development of the work reported in Guainazzi et al. 2006.}

\addkeyword{Galaxies:Active}
\addkeyword{Line: profiles}
\addkeyword{X-rays: galaxies}

\begin{document}
\maketitle

\section{Introduction}
The detection of a broadened and skewed Fe K$\alpha$ line in AGN  spectra is generally interpreted as an effect on X-ray photons due to the gravitational field of the black hole. Measuring the parameters of broad Fe lines provides therefore a diagnostic of the accretion disc structure and of the central object (see Fabian \& Miniutti 2005 for a review). 
Many publications on individual sources observed by the high throughput X-ray satellite
{\it XMM-Newton} invoke the presence of a relativistic disc Fe line (e.g. Wilms et al. 2001,
Fabian et al. 2002, Longinotti et al. 2003).
 Recent works on large samples of AGNs converged to say that the broad line is more common in low luminosity AGN (Nandra et al. 1997,2007, Streblyanska et al. 2005, Jimenez-Bail\`on et al. 2005, Guainazzi et al. 2006).
 Nonetheless, there is no agreement on the line parameters such as its intensity and equivalent width (EW).
Using a collection of 107 AGN from the {\it XMM-Newton} archive, Guainazzi et al. 2006 found a detection fraction of relativistic Fe line of 25\%. The mean EW was inferred to be $\sim$200~eV and the strongest lines were found in the sources with low 2-10~keV luminosity.
\section{This work: preliminary results}
\begin{figure}[!b]
  \includegraphics[width=\columnwidth]{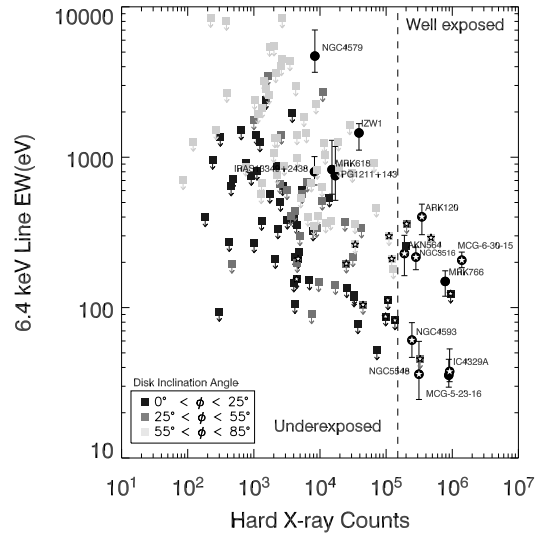}
  \caption{Equivalent width of the broad 6.4 Fe~K$\alpha$ line versus  the 2-10~keV counts. Filled circle: line detection at 5$\sigma$; filled squares: line upper limit at 97\% c.l. Stars mark sources in the flux-limited sample.}
  \label{fig:simple}
\end{figure}
We have expanded the work by Guainazzi  et al. (2006) by refining the baseline model and including more sources. 
The final sample is made by 157 type 1 radio-quiet AGN with N$_H$$<$10$^{22.5}$ cm$^{-2}$ (see Bianchi et al. 2007 for more details on the sample).  
The assumed baseline model is made by the following spectral components: 
primary X-ray power law,  Compton  reflection originated in the torus, 4 narrow emission lines with fixed energies corresponding to K$\alpha$ transitions from Fe I, XXV, XXVI and K$\beta$ from Fe I.  A Gaussian line of  50~eV width is included to fit the Compton Shoulder. The intensities of the Fe~K$\beta$ and of the Compton Shoulder are tied to the Fe~K$\alpha$ flux.
Absorption from ionized gas is also included to take into account any spectral curvature at 6-7~keV which may be induced by warm absorbing gas along the line of sight as for NGC 3783 (Reeves et al. 2004).
The {\it kyrline} model (Dov{\v c}iak et al. 2004) was adopted for fitting the relativistic disc line.
The presence of an (additional) reflection component arising from the disc and thus gravitationally blurred, 
 was extensively tested but it could not be distinguished from the torus reflection due to the limited  {\it XMM-Newton}  bandpass.
A single reflection component  is included in the final baseline model. 
When considering the whole sample, the broad line detection fraction is found to be of the order of 10\% for a significance threshold of 5$\sigma$ (Fig.1).
  However, a flux-limited (F$_{2-10}$$>$1.8$\times$10$^{-11}$ ergs~cm$^{-2}$~s$^{-1}$) sub-sample of 22 sources has been defined with the aim 
of performing a more thorough statistical analysis. The detection 
fraction in this subset rises up to 33\%.
 As pointed out by Guainazzi 2006, broad lines are mostly detected in well-exposed sources, i.e. in spectra  with a  number  of 2-10~keV X-ray counts greater than 1.5$\times$10$^5$~counts (Fig.1). 
These (still preliminary) results will be reported in  de la Calle et al. in prep.
\begin{figure*}[!t]
\includegraphics[width=6cm,height=6cm]{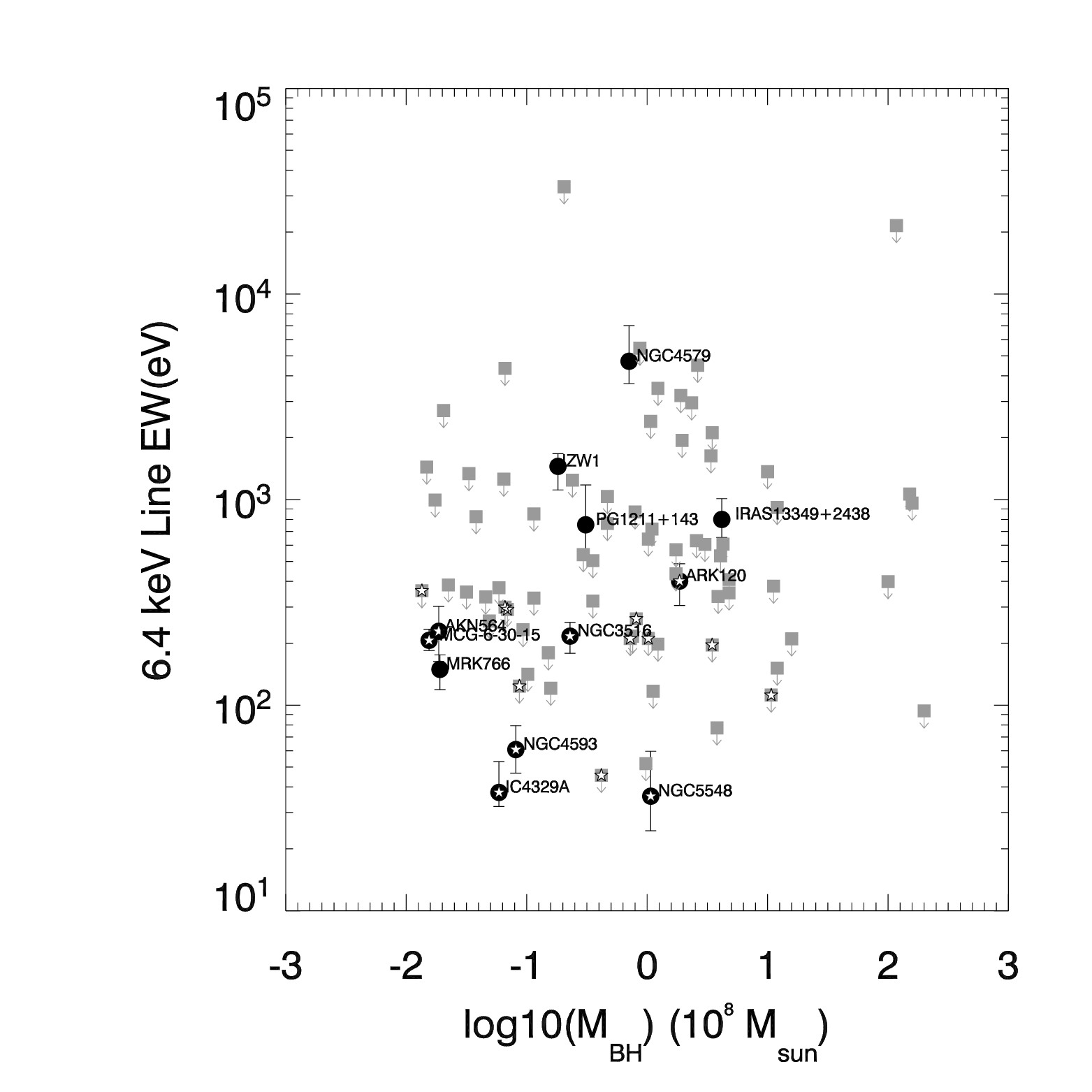}%
\includegraphics[width=6cm,height=6cm]{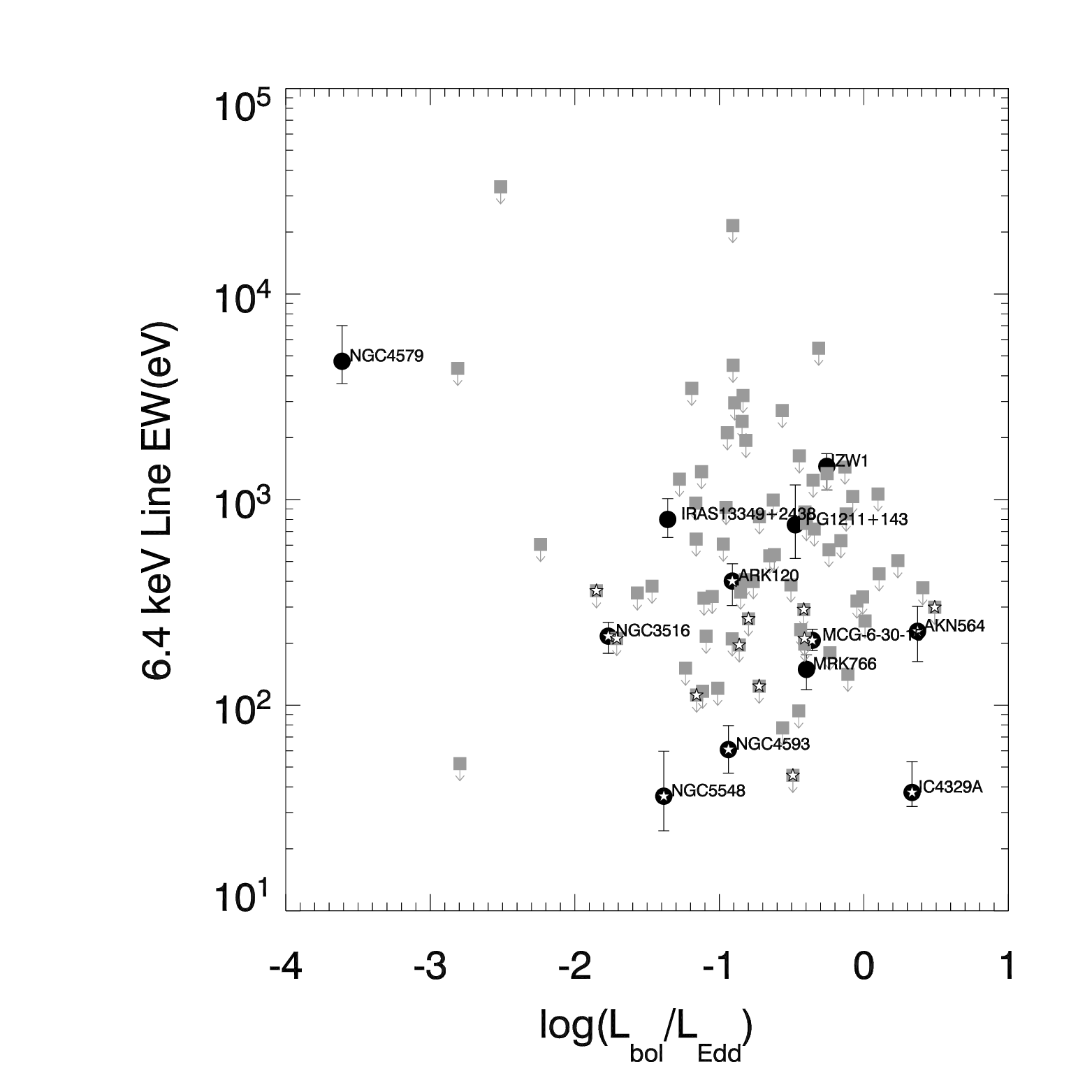}
\includegraphics[width=6cm,height=6cm]{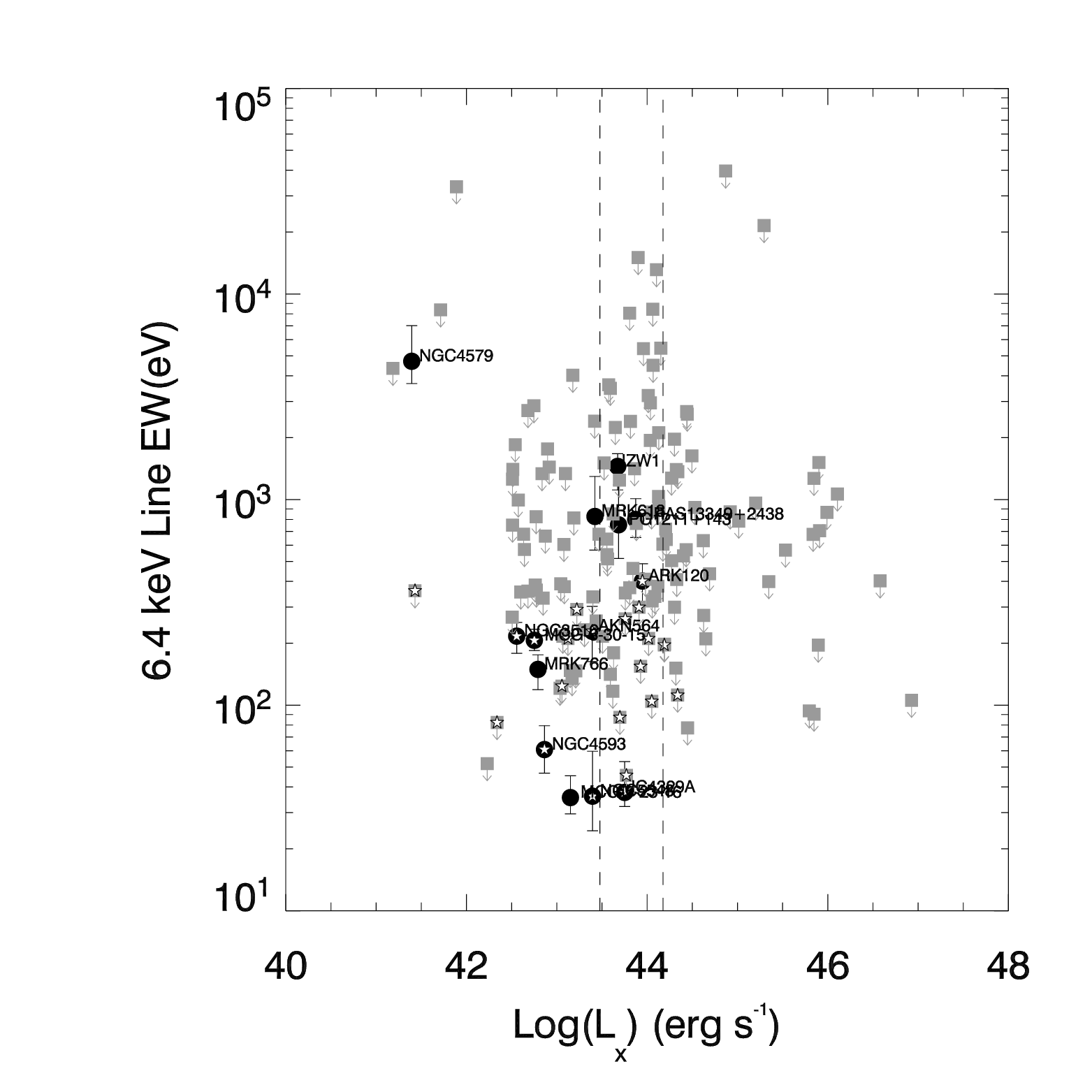}
\caption{Trend of the Fe K$\alpha$ EW versus physical properties of the sources.
Black circles mark the detections, grey squares mark the upper limits, stars mark sources in the flux limited sample. From left to right:EW versus black hole mass and EW versus accretion rate (for a subset of the sample); EW versus 2-10~keV luminosity (all sources in the sample, the dashed lines mark the same luminosity bins as Fig.3). 
No correlation is evident in any of the figures.}
\label{fig:widefig1}
\end{figure*}
For 65\% of the objects, measurements of H$\beta$ FWHM are available, 
allowing us to investigate if any relation holds between the presence of broad Fe line and the properties of the system such as black hole mass and accretion rate.
No apparent correlation seems to exist in any of the cases (Fig.2). However,  any circumstantiated conclusion should be made after performing the proper statistical tests
for which the results are not yet available.
Fig.~2 contains also the correlation of the broad line EW versus the X-ray luminosity (third panel): unlike the tight correlation found for the {\it narrow} Fe line EW and the X-ray luminosity (Bianchi et al, 2007),  no particular trend can be derived for the broad component.
This issue is examined in the following section. 
\section{Stacked spectra}
 To gather information on the spectra of under-exposed sources, 
the  spectral ratios of sources with no detection of broad line have been stacked together.
The ratio plots employed in the stacking were obtained assuming the continuum (with no emission features) 
as a baseline model.
Fig.~3a  shows the stacked plots for {\it all} the sources with upper limits. 
The prominent peak at 6.4 keV is produced by the narrow Fe~K$\alpha$ and the residuals at 
6.5-7.2~keV are likely to be due to the contribution of ionized Fe lines and possibly of Fe~K$\beta$.
There is no significant evidence for a  broad line profile  in the stacked spectra ratio, the data points below 6.4 keV are statistically consistent with the fitted continuum model.
Nonetheless, this ratio  was obtained  by ``mixing" sources with different X-ray luminosities.
To disentangle this effect the sources have been splitted in  3 luminosity groups (almost) equally populated.   
By comparing the profile of the stacked ratios to the theoretical  broad line profile (blue curve in the plots), 
 it can be concluded as a qualitative estimate that the line intensity never gets higher than 50~eV and it seems 
to faint with increasing luminosity.
More details will be presented in a forthcoming paper (Longinotti et al. in prep.)
\begin{figure*}[!t]
  \makebox[0pt][l]{\textbf{a}}%
  \hspace*{\columnwidth}\hspace*{\columnsep}%
  \textbf{b}\\[-0.7\baselineskip]
  \parbox[t]{\textwidth}{%
 \includegraphics[width=\columnwidth,height=6cm]{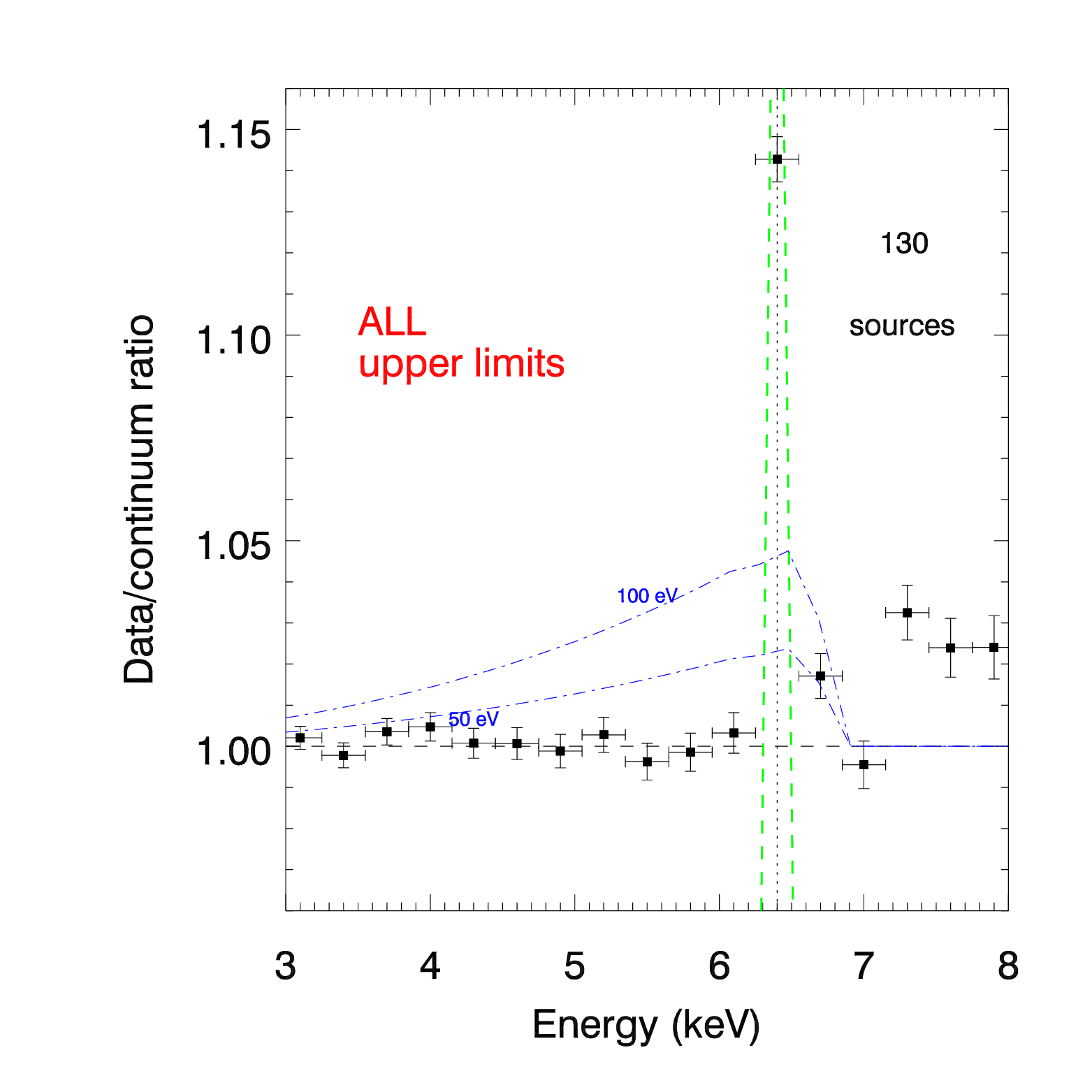}
     \includegraphics[width=\columnwidth,height=6cm]{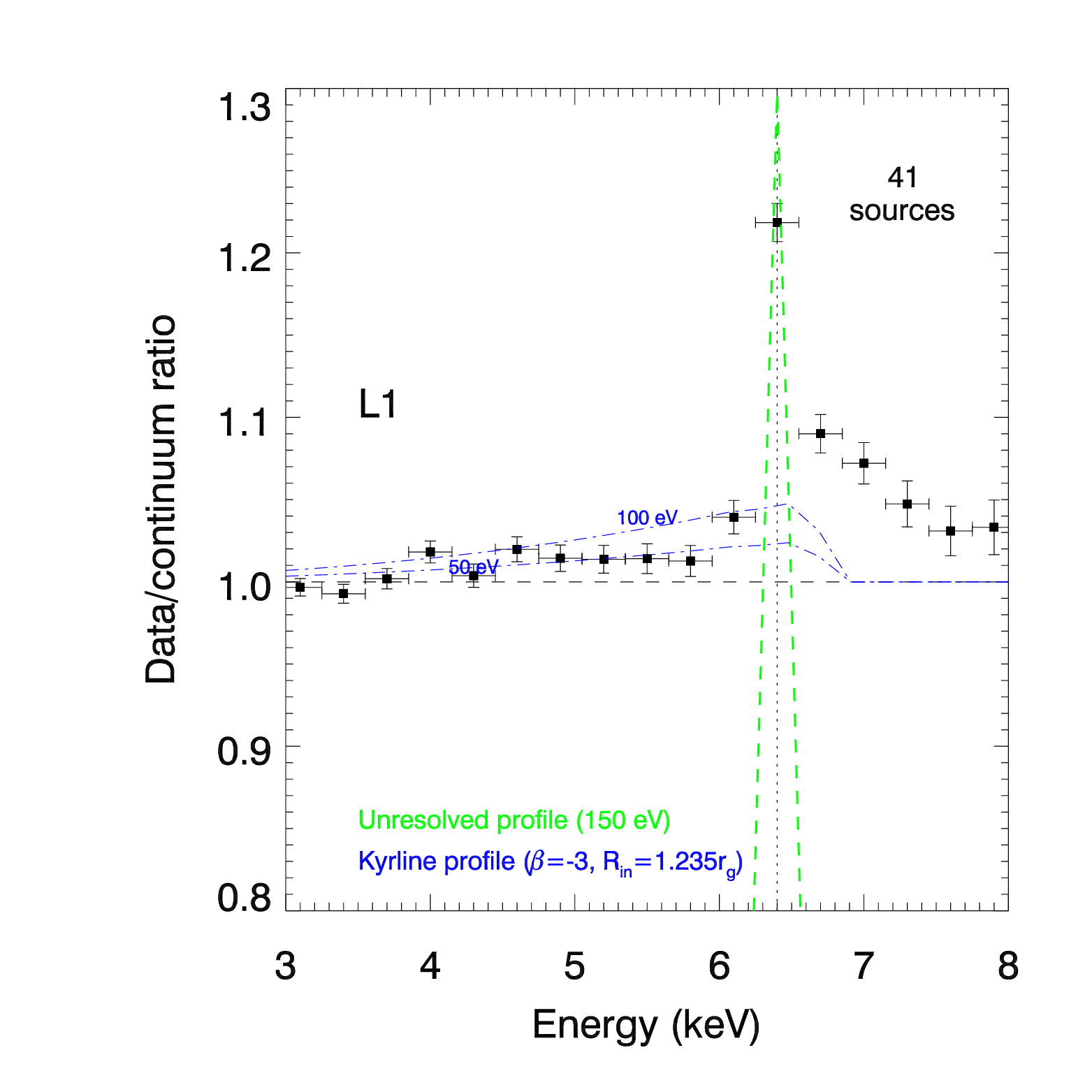}
 \makebox[0pt][l]{\textbf{c}}%
  \hspace*{\columnwidth}\hspace*{\columnsep}%
  \textbf{d}\\[-0.7\baselineskip]
 \includegraphics[width=\columnwidth,height=6cm]{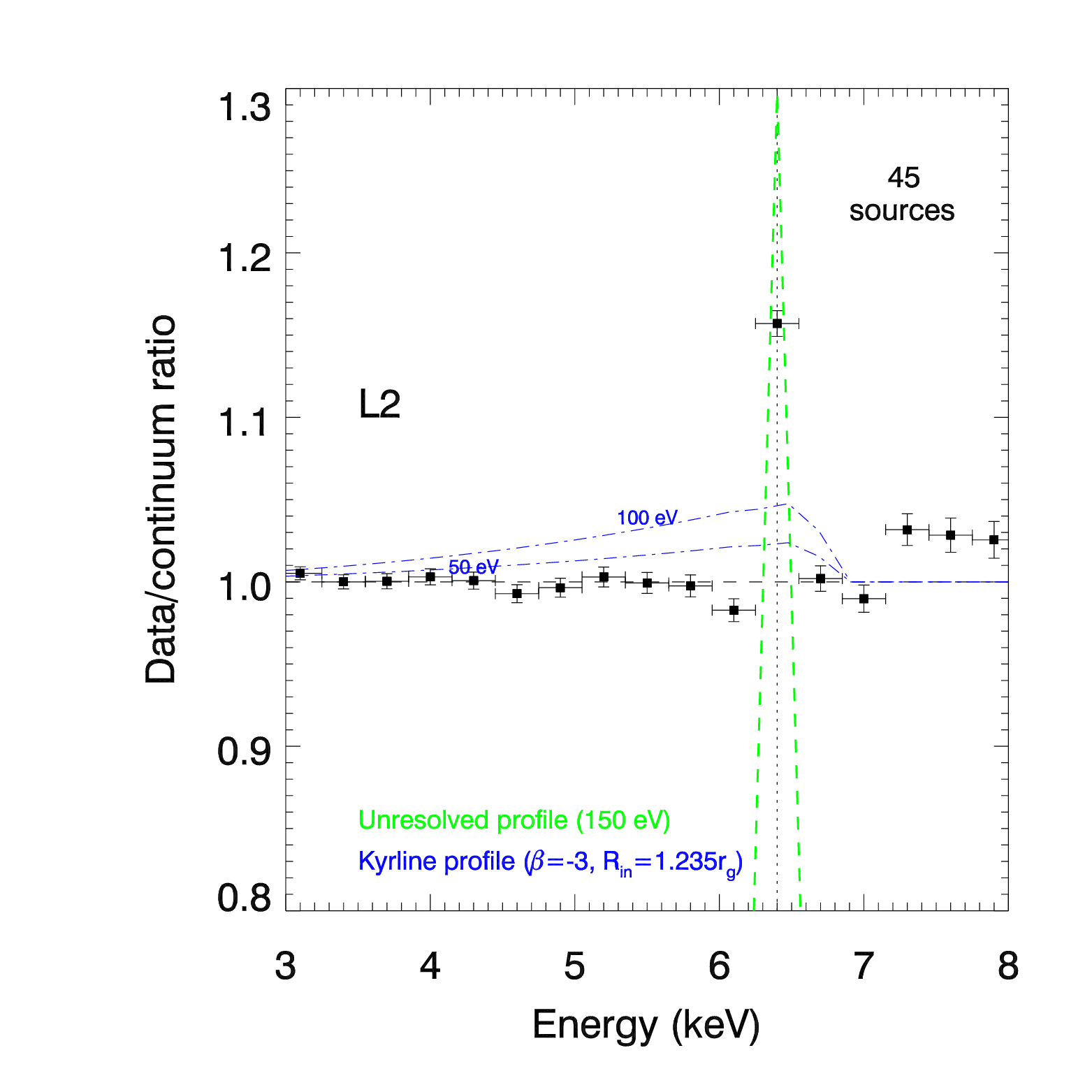}
 \hfill%
     \includegraphics[width=\columnwidth,height=6cm]{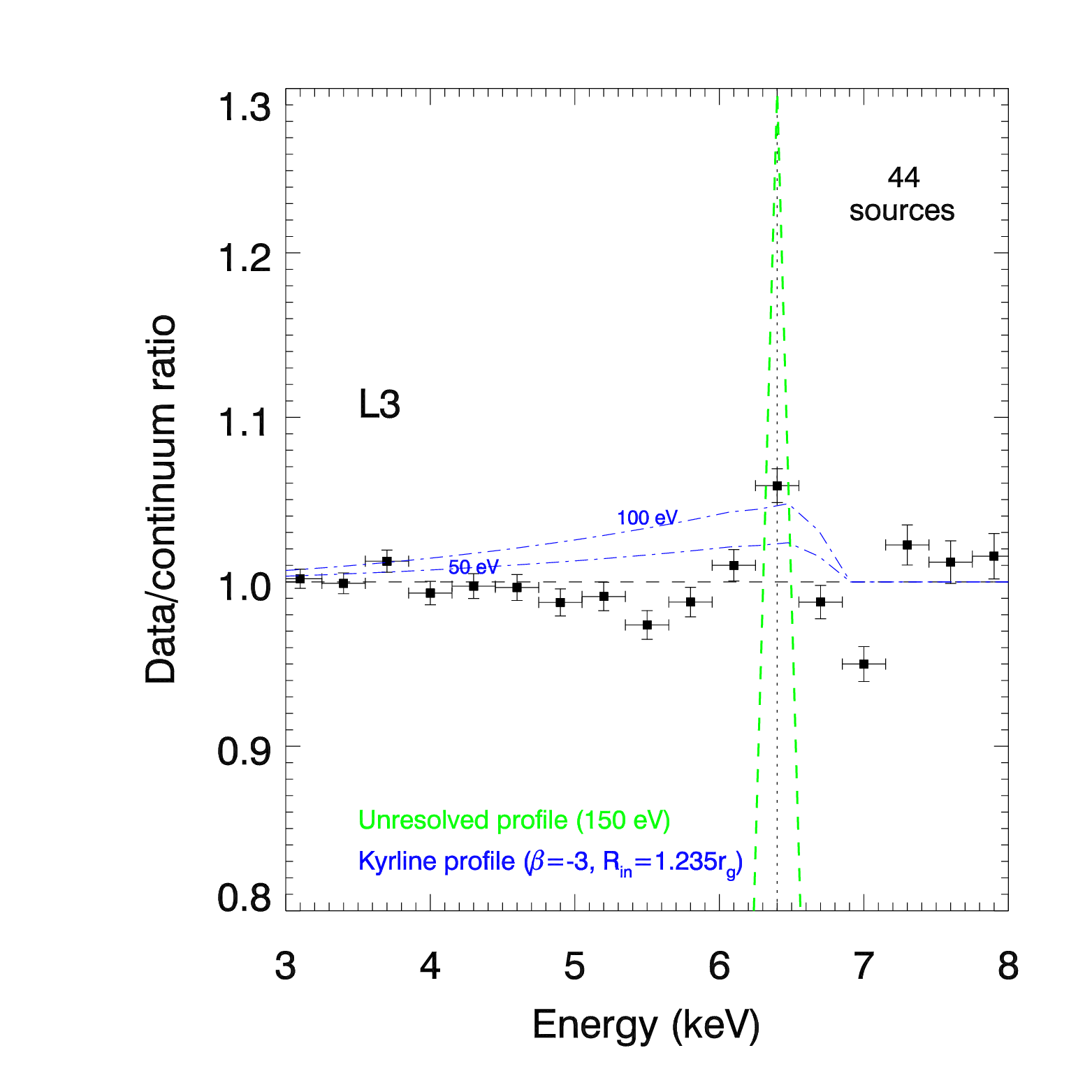}
     }
  \caption{(a) Stacking of all upper limits; (b,c,d,)  Iron line stacked residuals in the following hard X-ray luminosity groups (in ergs/s): (b) Lx$<$ 3$\times$10$^{43}$; (c) 3$\times$10$^{43}$ $<$ Lx $<$ 1.5$\times$10$^{44}$; (d) Lx $>$ 1.5$\times$10$^{44}$. The green and the blue profile represent the theoretical models for the narrow and the broad Fe line.}
\end{figure*} 

\section*{Acknowledgements}
Ignacio de la Calle would like to acknowledge support by the Torres Quevedo
fellowship from the Ministerio de Educaci\'on y Ciencia Espa\~nol and INSA.


\begin{thebibliography}
\bibitem[Bianchi et al.(2007)]{2007A&A...467L..19B} Bianchi, S., Guainazzi, 
M., Matt, G., \& Fonseca Bonilla, N.\ 2007, \aap, 467, L19
\bibitem[Dov{\v c}iak et al.(2004)]{2004ApJS..153..205D} Dov{\v c}iak, M., 
Karas, V., \& Yaqoob, T.\ 2004, \apjs, 153, 205
\bibitem[Fabian et al.(2002)]{2002MNRAS.335L...1F} Fabian, A.~C., et al.\ 
2002, \mnras, 335, L1
\bibitem[Fabian \& Miniutti(2005)]{2005astro.ph..7409F} Fabian, A.~C., \& 
Miniutti, G.\ 2005, arXiv: astro-ph/0507409 
\bibitem[Guainazzi et al.(2006)]{2006AN....327.1032G} Guainazzi, M., 
Bianchi, S., \& Dov{\v c}iak, M.\ 2006, Astronomische Nachrichten, 327, 
1032 
\bibitem[Jim{\'e}nez-Bail{\'o}n et al.(2005)]{2005A&A...435..449J} 
Jim{\'e}nez-Bail{\'o}n, E., Piconcelli, E., Guainazzi, M., Schartel, N., 
Rodr{\'{\i}}guez-Pascual, P.~M., \& Santos-Lle{\'o}, M.\ 2005, \aap, 435, 
449
\bibitem[Longinotti et al.(2003)]{2003A&A...410..471L} Longinotti, A.~L., 
Cappi, M., Nandra, K., Dadina, M., \& Pellegrini, S.\ 2003, \aap, 410, 471 
\bibitem[Nandra et al.(1997)]{1997ApJ...488L..91N} Nandra, K., George, 
I.~M., Mushotzky, R.~F., Turner, T.~J., \& Yaqoob, T.\ 1997, \apjl, 488, 
L91
\bibitem[Nandra et al.(2007)]{2007arXiv0708.1305N} Nandra, K., O'Neill, 
P.~M., George, I.~M., \& Reeves, J.~N.\ 2007, ArXiv e-prints, 708, 
arXiv:0708.1305
\bibitem[Reeves et al.(2004)]{2004ApJ...602..648R} Reeves, J.~N., Nandra, 
K., George, I.~M., Pounds, K.~A., Turner, T.~J., \& Yaqoob, T.\ 2004, \apj, 
602, 648 
\bibitem[Wilms et al.(2001)]{2001MNRAS.328L..27W} Wilms, J., Reynolds, 
C.~S., Begelman, M.~C., Reeves, J., Molendi, S., Staubert, R., \& 
Kendziorra, E.\ 2001, \mnras, 328, L27
\bibitem[Streblyanska et al.(2005)]{2005A&A...432..395S} Streblyanska, A., 
Hasinger, G., Finoguenov, A., Barcons, X., Mateos, S., \& Fabian, A.~C.\
2005, \aap, 432, 395
\end{thebibliography}
\end{document}